\documentclass[10pt]{iopart}
\usepackage{psfig}
\begin{document}

\title[{\boldmath $\phi$} puzzle in heavy-ion collisions at 2 AGeV]
{{\boldmath $\phi$} puzzle in heavy-ion collisions at 2 AGeV:\\
how many {\boldmath $K^-$} from {\boldmath $\phi$} 
decays?\footnote[1]{Supported by BMBF 06DR920,
06DR921, PROCOPE 9910330 and GSI.}}

\author{B K\"ampfer*, R Kotte*, C Hartnack\P \, 
and J Aichelin\P }

\address{*\
Forschungszentrum Rossendorf, D-01314 Dresden, PF 510119, Germany}
\address{\P\
SUBATECH and Ecole des Mines, 4, rue A. Kastler, B.P. 20722,
44307 Nantes, France}

\begin{abstract}
The preliminary experimental data on $\phi$ production in the
reaction Ni(1.93 AGeV) + Ni point to a puzzling high $\phi$
yield which can not be reproduced with present transport codes.
We survey the experimental situation and present prospects
for dedicated measurements of the $\phi$ multiplicities
with the $K^+ K^-$ and $e^+ e^-$ channels at HADES and FOPI.
\end{abstract}

\vspace*{-3mm}

\section{Introduction}

Within the SU(3) quark model, the $\phi$ meson has the composition
$\phi = \omega_8 \cos \Theta_V - \omega_1 \sin \Theta_V$
where 
$\omega_1 = (u \bar u + d \bar d +   s \bar s) / \sqrt{3}$
and
$\omega_8 = (u \bar u + d \bar d - 2 s \bar s) / \sqrt{6}$
are the singlet and octet representations. For a deviation
by $\Delta \Theta_V = 3.7^o$ from ideal mixing with 
$\Theta_V = 35.3^o$ \cite{PDG} the weight of the $s \bar s$
component in the $\phi$ is still 0.998, i.e.\ it consists
exceedingly of hidden strangeness. The branching ratio for
$\phi \to K^+ K^-$ is 0.491; via this decay channel the $\phi$ is accessible
e.g.\ in the FOPI and HADES detectors. 
The electromagnetic decay branching $\phi \to e^+ e^-$ is
$3.0 \times 10^{-4}$ thus allowing $\phi$ identification with HADES.
The narrow vacuum width of 4.43 MeV makes the $\phi$ meson an 
ideal object of in-medium hadron spectroscopy in strongly interacting
matter since any mass shift should become clearly visible,
in contrast to the wide $\rho$ meson.

Since the mass of the $\phi$ meson is only 32 MeV above the $K^+ K^-$
threshold a substantial coupling of the $\phi$ and $K^\pm$ dynamics
is to be expected. Due to this reason an understanding of the
role of the $\phi$ meson in intermediate-energy heavy-ion collisions
is highly desirable.  
It is the aim of this contribution to summarize the present
experimental situation and to consider prospects for future
measurements.

\section{Experimental situation}

The FOPI collaboration has collected about $4.7 \times 10^6$ central
events (10\% $\sigma_{\rm tot}$, $ \langle A_{\rm part} \rangle = 90$)
in the reaction Ni(1.93 AGeV) + Ni.\footnote{
There are also data for the reaction Ru(1.69 AGeV) + Ru, but 
not yet fully analyzed;
therefore we focus here on the reaction Ni(1.93 AGeV) + Ni.} 
The $\phi$ production is
deeply sub-threshold, as the threshold in pp collisions is at
2.59 GeV. Via identified $K^\pm$ pairs within the acceptance region
of the central drift chamber (CDC) original $\phi$ mesons are
reconstructed \cite{Norbert_1,Norbert_2}. The corresponding 
acceptance is centered around target rapidity and restricted by 
angular and momentum limits \cite{FOPI}.
Recently also $K^\pm$ candidates have been used to reconstruct
$\phi$ mesons within the HELITRON acceptance at midrapidity 
\cite{Roland_1}. The statistics is still too low to obtain any 
distribution. Rather, in both acceptance regions only the total 
numbers of $\phi$ mesons could be given. To accomplish an
extrapolation to full phase space, the assumption of an isotropic
distribution seems most natural as long as other information is
not at our disposal. This assumption must be supplemented
by additional information on parameters of the distribution.
Based on the analyses of $\pi^-$, $p$ and $d$ spectra \cite{Hong}
an effective temperature parameter of $T_{\rm eff} = 125$ MeV
was used in \cite{Norbert_1} to deduce a $\phi$ production
probability of $4 \times 10^{-4}$ which, with additional experimental
information, resulted in a $\phi / K^-$ ratio of about 0.1.
Later on, by improved analyses, both values were substantially
increased \cite{Norbert_2}. In \cite{Roland_2} the information
of the $\phi$ yields in the CDC and HELITRON acceptances were combined
and yielded an even larger $\phi$ multiplicity and a  $\phi / K^-$
ratio in the order of unity. The consistency of the CDC and
HELITRON data enforced a value of $T_{\rm eff} \approx 70$ MeV.
In estimating the $\phi / K^-$ ratio, $K^-$ measuments at KaoS
\cite{KaoS_1} and $K^- / K^+$ ratios together with $K^+$
multiplicities from FOPI \cite{Best} are employed and properly scaled
(for details cf \cite{Roland_2}).

\section{Theoretical situation}

\begin{figure}[b]
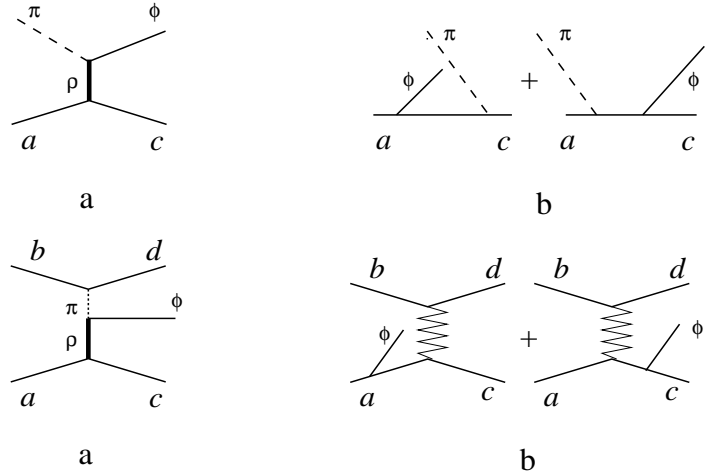

\hspace*{3cm}\psfig{file=figdia_1.epsi,width=9.3cm,angle=-0}
\vskip3mm
\hspace*{3cm}\psfig{file=figdia_2.epsi,width=9.3cm,angle=-0}
\caption{
Diagrams for the elementary $\phi$ production processes used
in \protect\cite{Titov}. 
Upper set: $\pi N \to N \phi$, 
lower set: $N N \to N N \phi$; $a, b$ ($c, d$) denote incoming
(outgoing) nucleons. 
Diagrams labeled by ''a'' are for
direct processes with a $\pi \rho \phi$ vertex, while
diagrams ''b'' include the $N N \phi$ coupling with poorly
known strength. The relative phases between diagrams ''a'' and
''b'' are important for interferences.}
\label{diagrams}
\end{figure}

First attempts to calculate the subthreshold $\phi$ production
in heavy-ion collisions have been performed by Ko and collaborators
within a BUU transport approach \cite{C.M.}. They included the channels
$N N$, $N \Delta$, $\Delta \Delta$, $\pi N$, $\pi \Delta$ and
$K \bar K$ $ \to \phi X$. The $K \bar K$ channel was found to
be small, while the $\pi N$ channel dominated. The parameterization 
used in \cite{C.M.} for the $pp \to pp \phi$ reaction underestimates 
the later on measured near-threshold data \cite{DISTO}. 
Therefore, we performed a new transport calculation by employing
the Nantes IQMD code \cite{IQMD} with newly parameterized elementary
cross sections. Within a one-boson exchange model,
in \cite{Titov} the $\phi$ cross sections $\pi N \to N \phi$ and 
$p p \to p p \phi$ have been adjusted to recent threshold-near data
\cite{DISTO}.
The corresponding diagrams are displayed in figure \ref{diagrams}.
The parametrization of the channels described above as well  as
the other $\phi$ production channels from 
$N N$, $N \Delta$, $\Delta \Delta$, $\pi N$, $\pi \Delta$
collisions are obtained by the following formulas:
\begin{eqnarray}
&&\sigma(pp \to pp \phi) = 5.78 \mu\mbox{b} \cdot 
\bigl( [\sqrt{s}-2m_n-m_\phi]/1{\rm GeV} \bigr)^{1.309} \\
&&\sigma(pn \to pn \phi) = 41.3 \mu\mbox{b} \cdot 
\bigl( [\sqrt{s}-2m_n-m_\phi]/1{\rm GeV} \bigr)^{1.438} \\
&&\sigma(\pi^+ n \to p \phi) = 101 \mu\mbox{b} \cdot 
\bigl( [\sqrt{s}-m_n-m_\phi]/1{\rm GeV} \bigr)^{0.466} \\
&& \sigma(\pi^+ n \to p \phi) = \sigma(\pi^- p \to n \phi) =
2 \sigma(\pi^0 p \to p \phi) = 2\sigma (\pi^0 n \to n \phi) \\
&& \sigma(nn \to nn \phi) = \sigma (pp \to pp \phi), \qquad
\sigma(\pi\Delta \to N \phi) = 0 \\
&&\sigma(N\Delta\to NN\phi) = 0.375 
\bigl( \sigma(pp \to pp \phi) + \sigma(pn \to pn \phi) \bigr) \\
&&\sigma(\Delta\Delta \to NN\phi) = 0.25 
\bigl( \sigma (pp \to pp \phi) + \sigma (pn \to pn \phi) \bigr) .
\end{eqnarray}

\begin{figure}[t]
 \centerline{
 \psfig{file=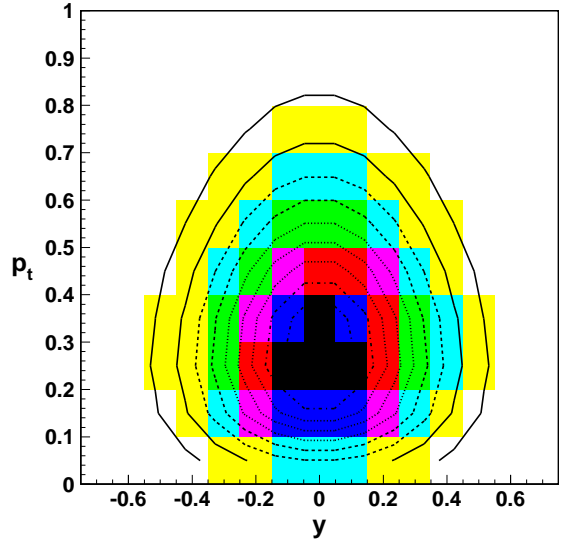,width=7.3cm,angle=-0}}
\caption{
Distribution $d^2N/dy dp_t$ of $\phi$ mesons in the reaction 
Ni(1.93 AGeV) + Ni according to the IQMD model. 
The lines represent the contour plot according to
equation (\protect\ref{isotropic_source}) with $T_{\rm eff} = 70$ MeV.
}
\label{sigma_IQMD}
\end{figure}

The resulting transverse momentum ($p_t$) vs rapidity ($y$) distribution 
for the reaction Ni(1.93 AGeV) + Ni
is displayed in figure \ref{sigma_IQMD}. Indeed, the distribution
looks very isotropic, 
as highlighted by a comparison with the parameterization  
\begin{eqnarray}
\frac{dN}{dm_\perp \, dy}
& = & 
{\cal N}
m_\perp^2 \cosh(y) \exp \left\{ 
-\frac{m_\perp \cosh(y)}{T_{\rm eff}} \right\}
\label{isotropic_source}
\end{eqnarray}
with $T_{\rm eff} \simeq 70$ MeV
($m_\perp=\sqrt{p_t^2+m_0^2}$ is the transverse mass).
While these calculations support the isotropy assumption made
in \cite{Norbert_1,Norbert_2,Roland_1,Roland_2} for a $4 \pi$
extrapolation, the total yields remain
below the preliminary experimental data.
This has inspired the authors of \cite{HW} to include in transport
model calculations further elementary reactions which contribute
to $\phi$ production. In particular, processes with incoming $\rho$
mesons are considered.
However, despite of an increase of the $\phi$ multiplicity for
the full phase space as well as for the CDC and HELITRON acceptance
regions separately, still the experimental values are underestimated.
This result calls for an explanation. In particular,
the preliminary experimental ratio $\phi / K^- \approx 1$ 
indicates that half of the
$K^-$ stem from $\phi$ decays. Otherwise, in many transport calculations
of the $K^-$ production (cf \cite{Nantes} for a recent study), 
the $\phi$ channel is not included at all;
inclusion of the surprizingly frequent $\phi$'s would then deliver too many
$K^-$.
If the ratio $\phi / K^- \approx 1$ would be experimentally confirmed,
then in previous experiments too few $K^-$ were seen. This appears as
a puzzling situation which is caused by the restricted and partially
disjunct phase space regions where the $\phi$ and $K^-$ are measured.

Notice in this context that HSD transport calculations for central 
Au + Au reactions \cite{Cassing}
give a ratio $\phi / K^- \approx 1$ at beam energy of about 1.5 AGeV,
while at 1.9 AGeV the ratio drops below unity.\\[3mm]

\begin{figure}[h]
\centerline{\psfig{file=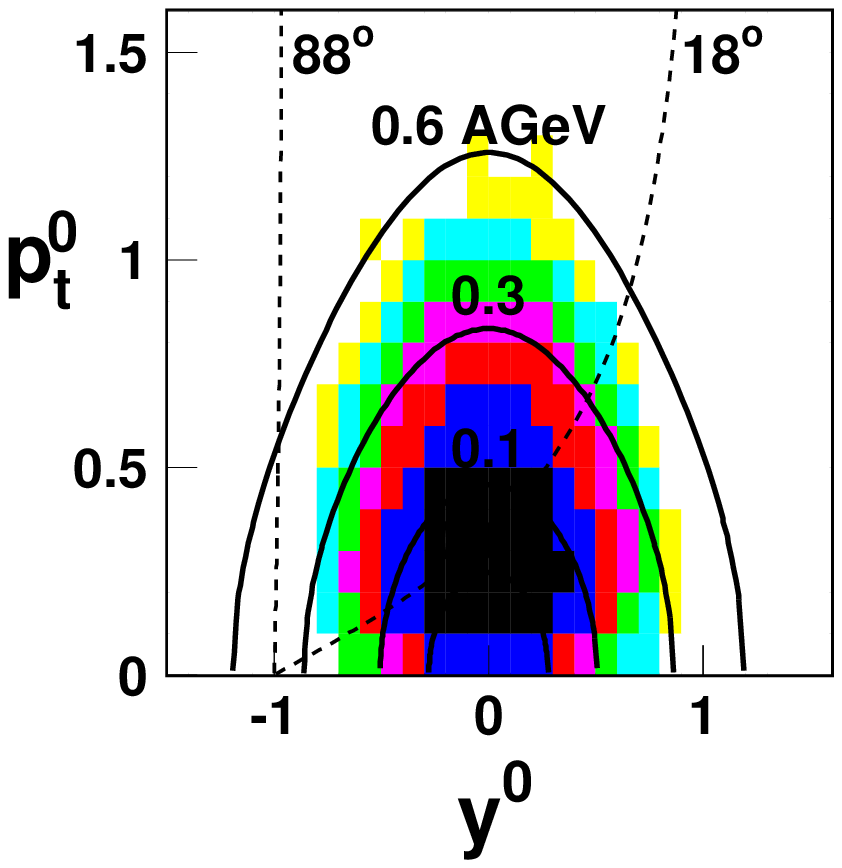,width=6.3cm,angle=-0}
\hfill
\psfig{file=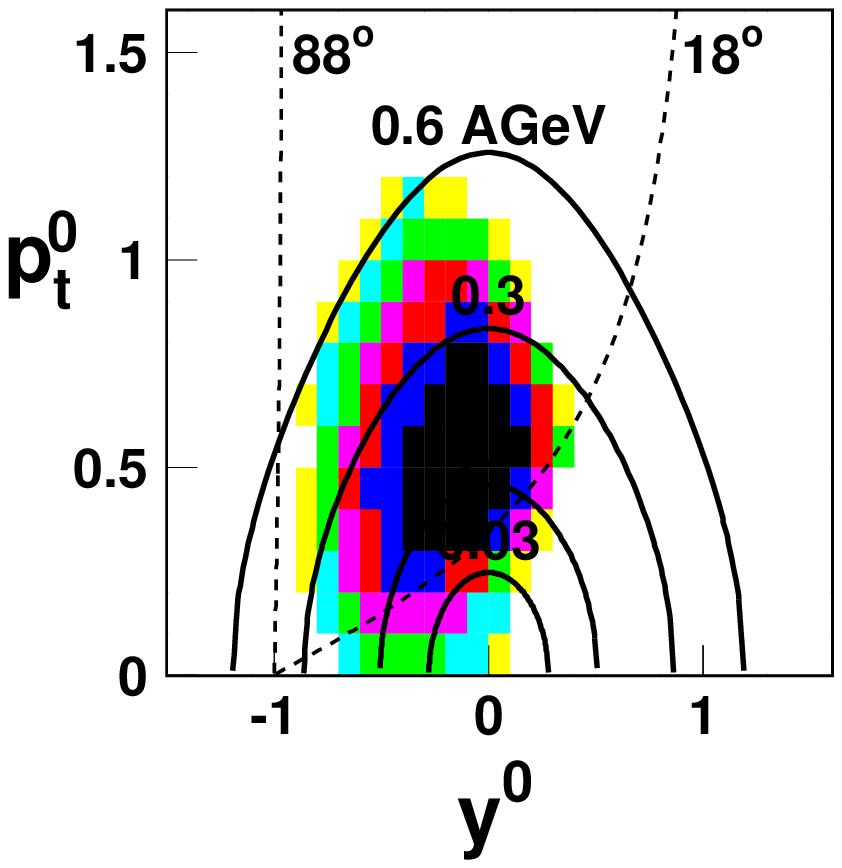,width=6.3cm,angle=-0}}

\vspace*{6mm}

\centerline{\psfig{file=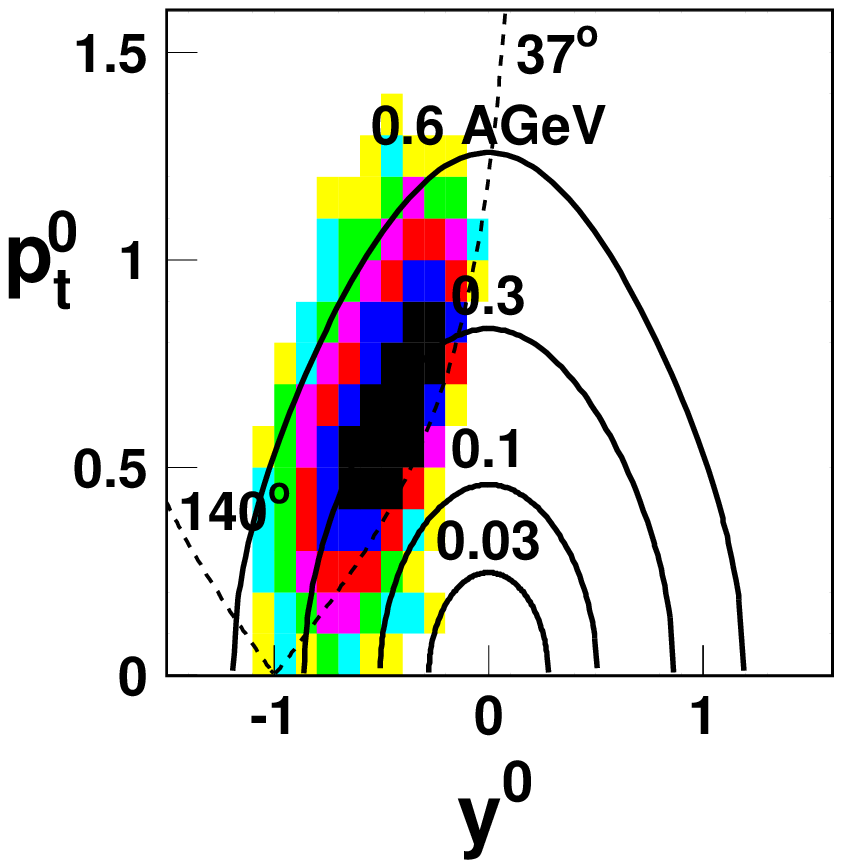,width=6.3cm,angle=-0}
\hfill
\psfig{file=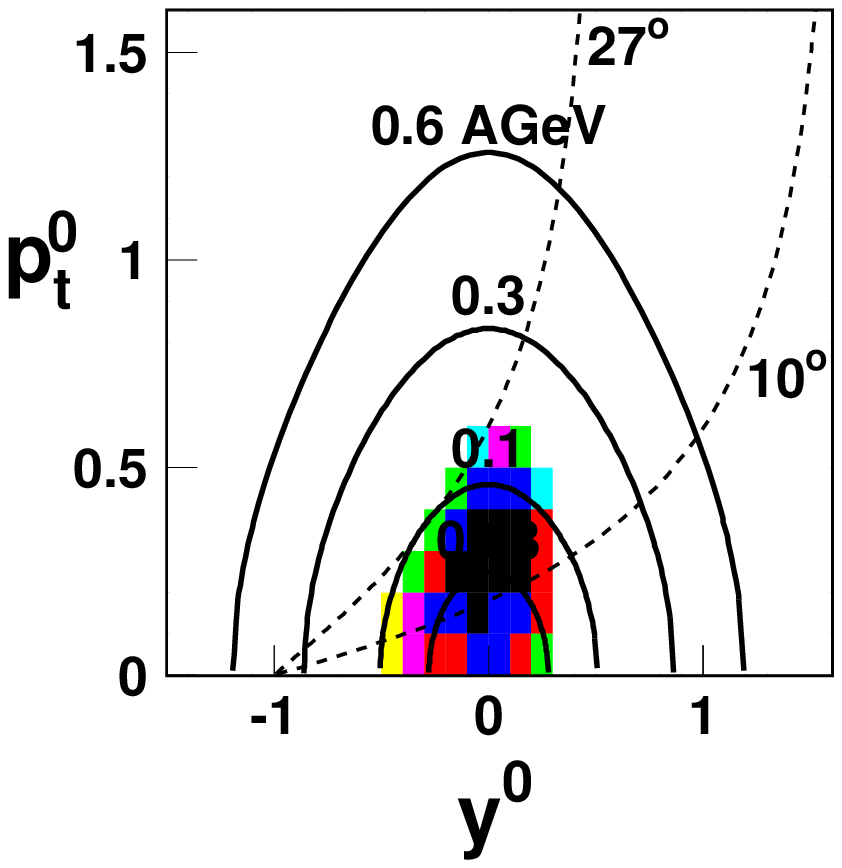,width=6.3cm,angle=-0}}
\caption{
Phase-space distribution $d^2N/dp_t dy$ 
of $\phi$ mesons within the acceptance regions of HADES 
(figures on top,
left panel: $e^+ e^-$ channel, 
right panel: $K^+ K^-$ channel)
and FOPI 
(figures on bottom, $K^+ K^-$ channel, 
left panel: $K^\pm$ in 
the acceptance of CDC with upgraded Plastic Barrel,
right panel: $K^\pm$ within HELITRON). 
Here, 
$p_t^0=(p_t/A)/(p_{proj}/A_{proj})_{cm}=
(\beta_t \gamma)/(\beta \gamma)^{proj}_{cm}$ and 
$y^0=(y/y_{proj})_{cm}=(y/y_{cm}-1)$
are the normalized transverse momentum and rapidity, respectively, 
and $A=m/m_p$ is the particle mass number. Both observables 
are related to the corresponding projectile quantities in the 
center-of-mass (c.m.) frame 
of the colliding nuclei (with $y_{cm}=0.904$ and 
$(\beta \gamma)^{proj}_{cm}=1.032$ for
2.0 $A\cdot$GeV beam energy). A logarithmic intensity 
scale is used.}
\label{HADES_FOPI}
\end{figure}

\section{Prospects of improved measurements}

To elucidate the maximum count rates of $\phi$ mesons in future
experiments we present here estimates 
for the symmetric reaction C(2.0 AGeV) + C which are based on 
(i) the isotropic distribution in equation (\ref{isotropic_source})
with $T_{\rm eff} = 70$ MeV, 
(ii) target thickness corresponding to 1\% interaction probability,
(iii) maximum beam intensities compatible with the data acquisition
times and taping rates for the detector systems HADES and FOPI, and
(iv) acceptance regions, efficiencies, and geometry of the detectors.
The corresponding details are described in \cite{Roland_bk}.
The normalization factor ${\cal N}$ in equation (\ref{isotropic_source})
corresponds to a $\phi$ production probability of
$P_\phi = (\sigma_\phi / \sigma_{K^-}) \times 
(\sigma_{K^-} / \sigma_{\rm tot})$, where the first term is 0.2
from \cite{E917_E866} and the second one equals $2 \times 10^{-4}$
from \cite{KaoS_2}.
Taking into account the corresponding branching ratios,
the maximum count rates of $\phi$ mesons per day are:
(i) 30 (HADES, $e^+ e^-$ channel, $18^o \le \Theta_e \le 88^o$),
(ii) 5 (HADES, $K^+ K^-$ channel, $44^o \le \Theta_K \le 88^o$,
and $p_K < 1$~GeV/c), 
(iii) 160 (HADES, $K^+ K^-$ channel, $18^o \le \Theta_K \le 88^o$ 
and $p_K < 1$ GeV/c),
(iv) 50 (FOPI, $K^+ K^-$ channel, detector 
combination of CDC and upgraded Barrel:  
$37^o \le \Theta_K \le 140^o$ and 0.1~GeV/c $<p_K < 1$~GeV/c, 
combination of HELITRON and Plastic Wall: 
$10^o \le \Theta_K \le 27^o$ and 
0.2~GeV/c $<p_K < 0.8$~GeV/c, as well as mixed combination
of CDC and HELITRON).
These numbers do not include any background estimates.
While (iii) looks very promising, it needs an upgrade 
since at present both the granularity and the 
time-of-flight resolution within the TOFino angular
region  $18^o \le \Theta_K \le 44^o$ are not sufficient 
for $K^\pm$ identification. 

At HADES the $e^+ e^-$ channel gives the most
complete phase space distribution ($\sim 40$\,\% geometrical
acceptance), 
as seen in the upper left panel of figure \ref{HADES_FOPI}.
Though the $K^\pm$ channel is accessible with an order 
of magnitude lower acceptance  
it will still allow for reliable reconstruction of the $\phi$
distribution and the corresponding total production
probability (see upper right panel of figure \ref{HADES_FOPI}).  
No momentum restriction is invoked in case of the leptonic 
decay while for the hadronic branch the $K^{\pm}$   
momenta are restricted to $p_K<1$~GeV/c for 
ensuring the particle identification. 
The corresponding cut squeezes the 
covered phase space distributions only marginally.

At FOPI the $\phi$ identification via the $K^\pm$ channel
within the CDC/Barrel acceptance gives access only to the target 
rapidity region (see lower left panel of figure 
\ref{HADES_FOPI}). The phase space coverage is enlarged hence  
allowing access to the region of highest phase-space density, i.e.  
at midrapidity and at small transverse momentum,  
when exploiting the HELITRON/Plastic Wall subdetector 
combination (see lower right panel of figure \ref{HADES_FOPI}). 
When combining the kaons of the HELITRON with the antikaons of the CDC 
even the intermediate region can be populated (not displayed). 
Further details of these simulations can be found 
in \cite{Roland_bk}.

\section{Summary}

In summary we refer to transport model calculations which 
underestimate the preliminary $\phi$ yields 
in individual phase space regions.
This together with a large ratio $\phi / K^- = {\cal O} (1)$   
points to the need of an improved understanding of the $\phi$
dynamics in intermediate-energy heavy-ion collisions.
We present a series of simulations to elucidate the prospects
for dedicated $\phi$ measurements in the $K^+ K^-$ and $e^+ e^-$
channels at already existing facilities at SIS/GSI Darmstadt.

\end{document}